# Roadside acoustic sensors to support vulnerable pedestrians via their smartphone

Masoomeh Khalili, Mehdi Ghatee, Mehdi Teimouri, and Mohammad Mahdi Bejani

***Abstract*—We propose a new warning system based on smartphones that evaluates the risk of motor vehicle for vulnerable pedestrian (VP). The acoustic sensors are embedded in roadside to receive vehicles sounds and they are classified into heavy vehicle, light vehicle with low speed, light vehicle with high speed, and no vehicle classes. For this aim, we extract new features by Mel-frequency Cepstrum Coefficients (MFCC) and Linear Predictive Coefficients (LPC) algorithms. We use different classification algorithms and show that MLP neural network achieves at least 96.77% in accuracy criterion. To install this system, directional microphones are embedded on roadside and the risk is classified there. Then, for every microphone, a danger area is defined and the warning alarms have been sent to every VPs' smartphones covered in this danger area.**

## I. Introduction

Smartphones are the best options for safety purposes in the recent years [1], [2], [3], [4]. These instruments are used for pedestrian safety in the current paper. Pedestrian safety is an important challenge in the world, but they like to carry least equipment to protect themselves. Sometimes, infrared sensor, camera, computer vision and wireless networks were used to recognize the environment [5]. On the other hand, to improve safety, it is necessary to classify risk of different situations. For this aim, different sensors are used to monitor and to classify the roads and vehicles. For example, intrusive sensors such as inductive loops [6], magnetometers, micro loop probes, pneumatic road tubes and piezoelectric cables [7] are commonly used. In [8], the performance of these sensors has been compared. Also as classification perspective, usually, vehicles are classified without pedestrian's properties. For example, Nooralahiyan et al. [9] proposed a vehicle identification method using Linear Predictive Coding (LPC) based on acoustic sound source of moving vehicle. Wavelet packet algorithm for moving vehicle classification has been proposed in [10]. In [7] by using ANN, vehicles are classified into heavy vehicle, medium vehicle, light vehicle, and horns classes with accuracy equal to 67.4%. In addition, in [11], a vehicle sound classification system is presented when low pass filtering was performed that just classifies into six models of cars that is not helpful for VP. In [12], a quadratic discriminant analysis was used to classify audio signals of passing vehicles and its accuracy is 80%. In [13], vehicle sound was classified by using Probabilistic Neural Network (PNN). In [14] an acoustic hazard detection system was developed for pedestrians with obscured hearing that was not mentioned anything about accuracy of classification and vehicle direction detection. Generally, in these works, we face with some drawbacks. They used some unusual and expensive equipment for VP. Some extra information from vehicle driver and pedestrian are needed and just a communication between a single car and a single pedestrian is supported. In addition, there is a high cost for installation and maintenance for sensors such as camera. Furthermore, it is not enough clarity in helping pedestrians with respect to vehicle direction when we use acoustic sensors.

Based on these works, in this paper we focus on VP including pedestrians who use voice players or work with mobile applications and their attention to risky patterns decreases. In addition, cyclists and wacky pedestrian are considered as VP in this paper. For this kind of VPs, we propose a new warning system based on received sounds in environment. The main contribution of this paper is to use acoustic data with low installation and maintenance costs, light processing, and good performance in different weather conditions such as sunny, rainy and foggy, good performance in the dark and not enough light. To use acoustic sensors for risk evaluation of vehicles in road, we just assume that VP has a smartphone with internet connection and we do not need extra information from vehicle driver such as location and speed and age. We prove that our system provides high accuracy in vehicle sound classification and we consider vehicle direction to reduce false alarms. Furthermore, environment includes many sources of noises and so it is impossible to consider all of the situations. In this paper, we consider some noise effects to improve the system performance.

- Masoomeh Khalili is an MSc student with Department of Computer Science of Amirkabir University of Technology, Tehran, Iran.
- Mehdi Ghatee is an Associate Professor with Department of Computer Science, Amirkabir University of Technology, Tehran, Iran. Email Address: ghatee@aut.ac.ir, URL: www.aut.ac.ir/ghatee
- Mehdi Teimouri is an Assistant Professor with Faculty of New Sciences and Technologies, University of Tehran, Tehran, Iran.
- Mohammad Mahdi Bejani is a PhD student with Department of Computer Science of Amirkabir University of Technology, Tehran, Iran.

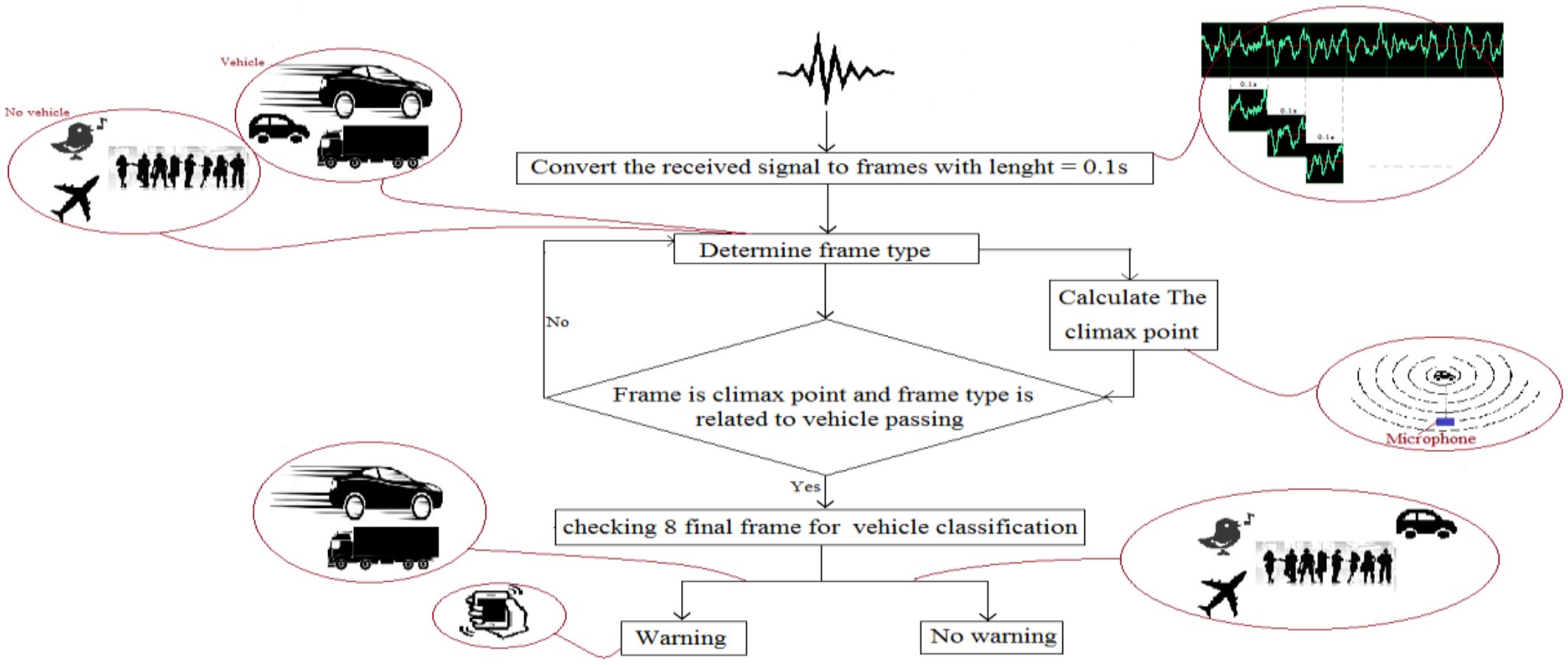

Fig. 1. Warning system by using vehicle classification for VPs

## II. Proposed System

In this section, we propose a warning system for VP when the road is straight without intersection. The system includes two modules: Sensors and Decision-making. Fig. 1 shows a graphical abstract of the proposed system. In what follows, we explain the details.

### *A. Sensors*

Fig. 2 shows the embedded equipment in the roadside and danger area. Distance between two successive processors is 25m. Each processor has a directional microphone (with 3m height) to detect direction in two-way streets when passing a vehicle. Danger area is also defined as a rectangular with 25m length and each processor corresponds to a special danger area that is shown in Fig. 2. These distances are determined experimentally. On the other hand, if the maximum speed to detect is assumed 75 km/h and since we need at least 3 seconds to warn the VP [15] [16], so a vehicle can pass 75m in 3.6 second. Therefore, from 75m ago, we have enough time to warn VP and according to danger area length, the minimum time and the maximum time to warn VP is 3.6 and 4.8 seconds, respectively.

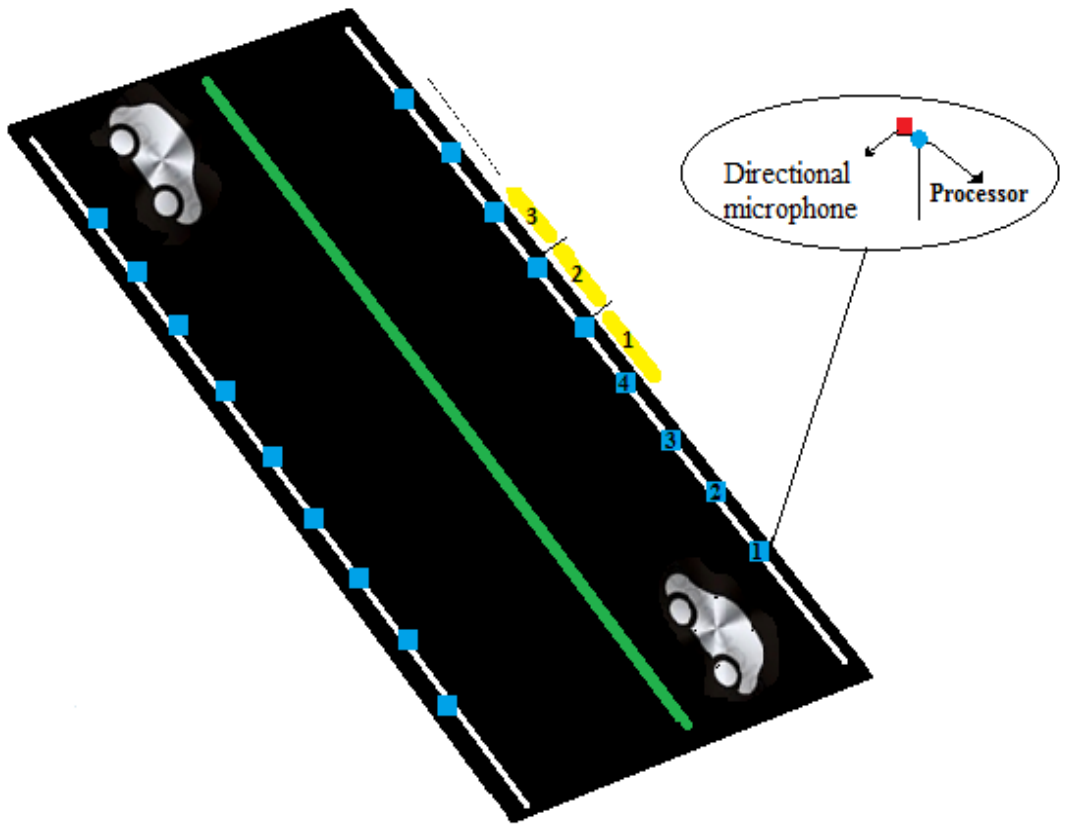

Fig. 2. Roadside equipment of the proposed warning system (Blue circle: processor; Red square: directional microphone; Yellow color: danger area;)

### *B. Decision making*

When a vehicle approaches to a processor, microphone receives the vehicle sound. System detects vehicle direction and classifies the vehicles into in four classes: heavy vehicle (H), light vehicle with low speed (LL), light vehicle with high speed (LH) and no vehicle (NV). NV includes birds' sound, airplanes, and crowds of people. By the aid of directional microphones, vehicle direction can be detected simply. In this research, we use Cardioid directional microphone that provides great sensitivity at the front, only partially at the sides, and little at the back. By using this microphone, the proposed system can detect the right direction of vehicle. In this system, after the sound signal is received, it converts to the frames with 0.1 second length and for each frame; the following phases should be done.

*Phase 1: Determining the frame type*

This phase includes three steps. In the first step, features of Table I are extracted. To determine the first five features, we use Fast Fourier Transform (FFT). Fig. 3 shows FFT of a vehicle sound. Let $x_1, \ldots, x_{N-1}$ be complex numbers, Equation (1) shows FFT sequence.

$$x_k = \sum_{n=0}^{N-1} x_n e^{-i2\pi kn/N} \qquad k = 0.1.\ldots.N-1 \qquad (1)$$

The first and the second features reach form signal power in the first and the second half, presented in Fig. 3. The third and the fourth features determine as the frequency of the highest value in the first and second half of the FFT. The fifth feature is shown in Fig. 3 with yellow circle as a point that is similar in both sides of half parts of FFT. The sixth and seventh features are corresponded to the outputs of MFCC and LPC algorithms. MFCC is one of the most popular algorithms in speech processing [17]. The details of this algorithm is given in Fig. 4. In addition, LPC algorithm provides reliable, robust and accurate estimation of speech parameters. In Fig. 5, LPC steps are shown [18].

In the second step of phase 1, we implement a feature selection method namely Principal Component Analysis (PCA) which is a standard tool in modern data analysis for extracting relevant information from confusing data sets. The final step is classification. In this step, we test four classification algorithms including Multi-Layer Perceptron (MLP), K-Nearest Neighbor's algorithm (KNN), Naive Bayes (NB) and Decision Tree (DT). We compared the results of these classification algorithms in the next section. After this step, every frame type is classified.

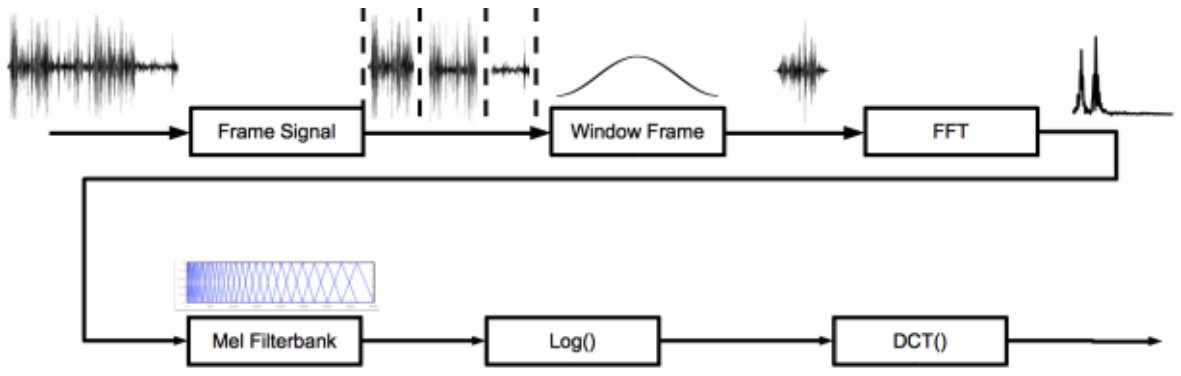

Fig. 4. MFCC steps adopted from [14].

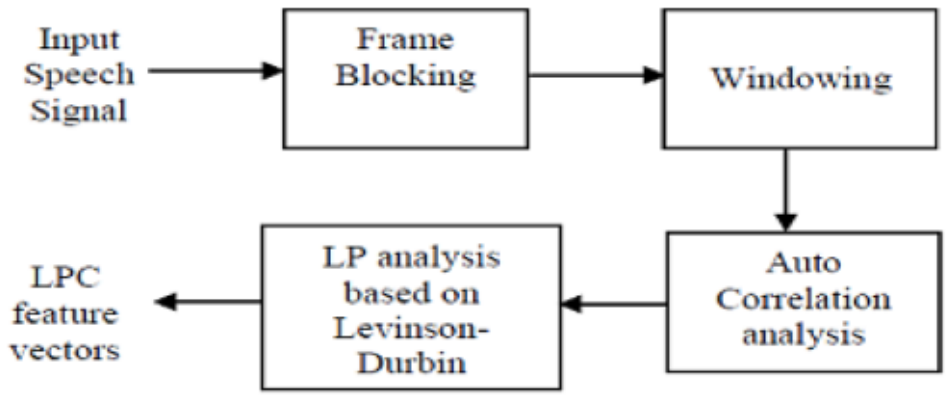

Fig. 5. LPC steps adopted from [15].

*Phase 2: Calculating the climax point*

Climax point is a point that a vehicle is crossing from the nearest point to acoustic sensor that have highest frequency. For finding the climax point, we use Doppler Effect. Doppler Effect is a very important physical phenomenon with a great variety of applications [19]. The Doppler Effect consists of a change in the frequency received by the receptor when the source moves relative to it, that is, the frequency increases when the receptor approaches the source and decreases when the receptor moves away from the source [19]. This means that when a vehicle crosses from the nearest point to a processor, receives the highest frequency.

*Phase 3: Checking eight final frames*

At the end of decision making part, if the climax point is not relevant to frames of NV type, then the third phase begins. In this phase, we use eight last frames up to climax point, to decide about the received sound. We use eight last frames experimentally. If the number of frames with similar type is the greatest, then this frame type is presented as the received sound type.

TABLE I

FEATURES THAT USE IN PROPOSED SYSTEM

| | Features |
|---|---|
| 1 | $P_1$: Power of the first half of FFT result |
| 2 | $P_2$: Power of the second half of FFT result |
| 3 | $F_1$ : Frequency of the highest power in the first half of the FFT |
| 4 | $F_2$: Frequency of the highest power in the second half of the FFT |
| 5 | The highest value |
| 6 | MFCC |
| 7 | LPC |

## C. Warning

When the direction is detected and the vehicles are classified, if there is a risk made by a heavy vehicle or light vehicle with high speed, then the proposed system sends a warning massage on VPs' smartphones that are located in the corresponding danger area. In this system, we need to find the location of VP by the aid of Global Positioning System (GPS). When VP need to became aware of a risky vehicle, which is approaching, it is necessary to turn on GPS of her/his smartphone. Then the roadside processor can send warning to VP through an application installed on smartphone. In addition, the roadside processor just sends the messages to smartphones located in its danger area. According to [20], the maximum acceptable error in warning systems is one meter and in [21], the best error of GPS was reported between 0.01 and 1 meter. Thus, the usage of GPS for positioning is acceptable in the proposed system.

# III. EXPERIMENTAL RESULT

In Table II the best results about the assistant systems for pedestrians are presented. In the final column of this table, the preferences of our proposed system compared with the previous systems are presented. Also in Table III, the best results on vehicles sound classification is shown. In addition, we present the advantages of the proposed system compared with these vehicles sound classification is shown. In addition, we present systems. However, to compare the numerical results of the proposed system, we need some

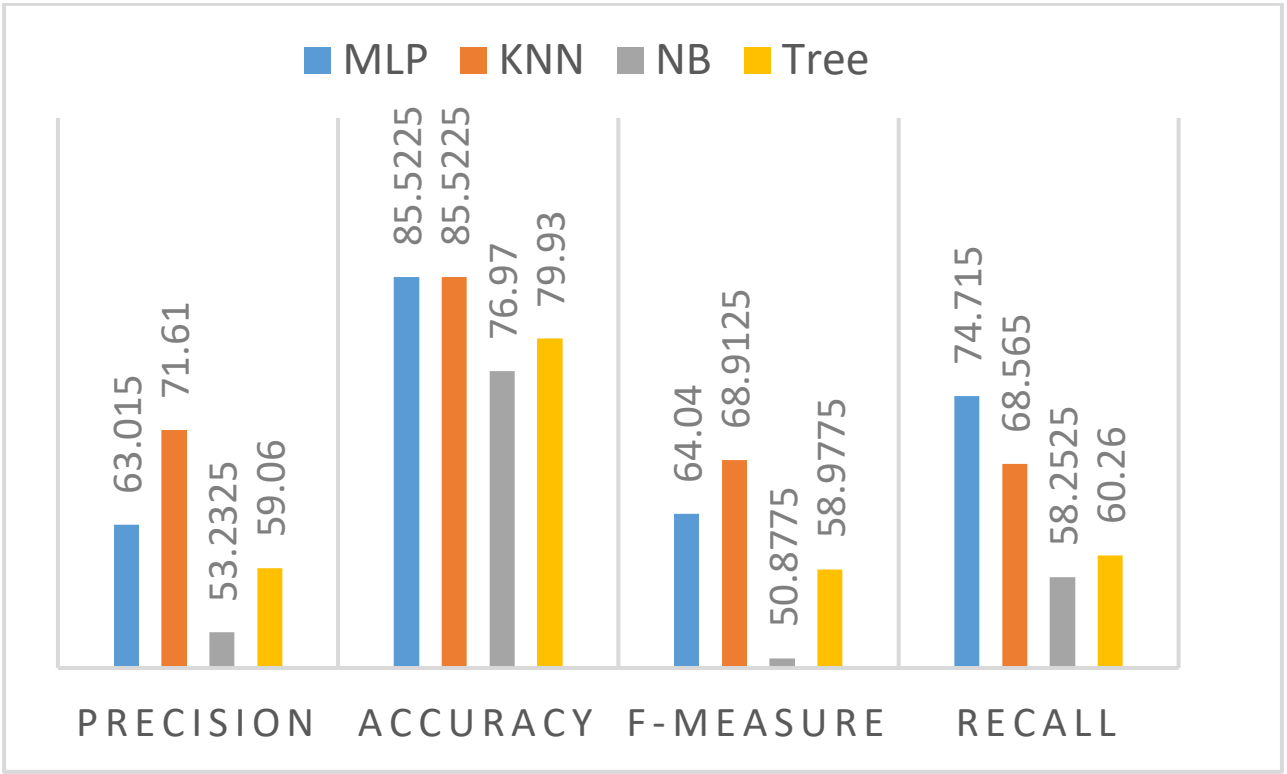

Fig. 6. Comparison between different classification algorithms using the first five features of Table I

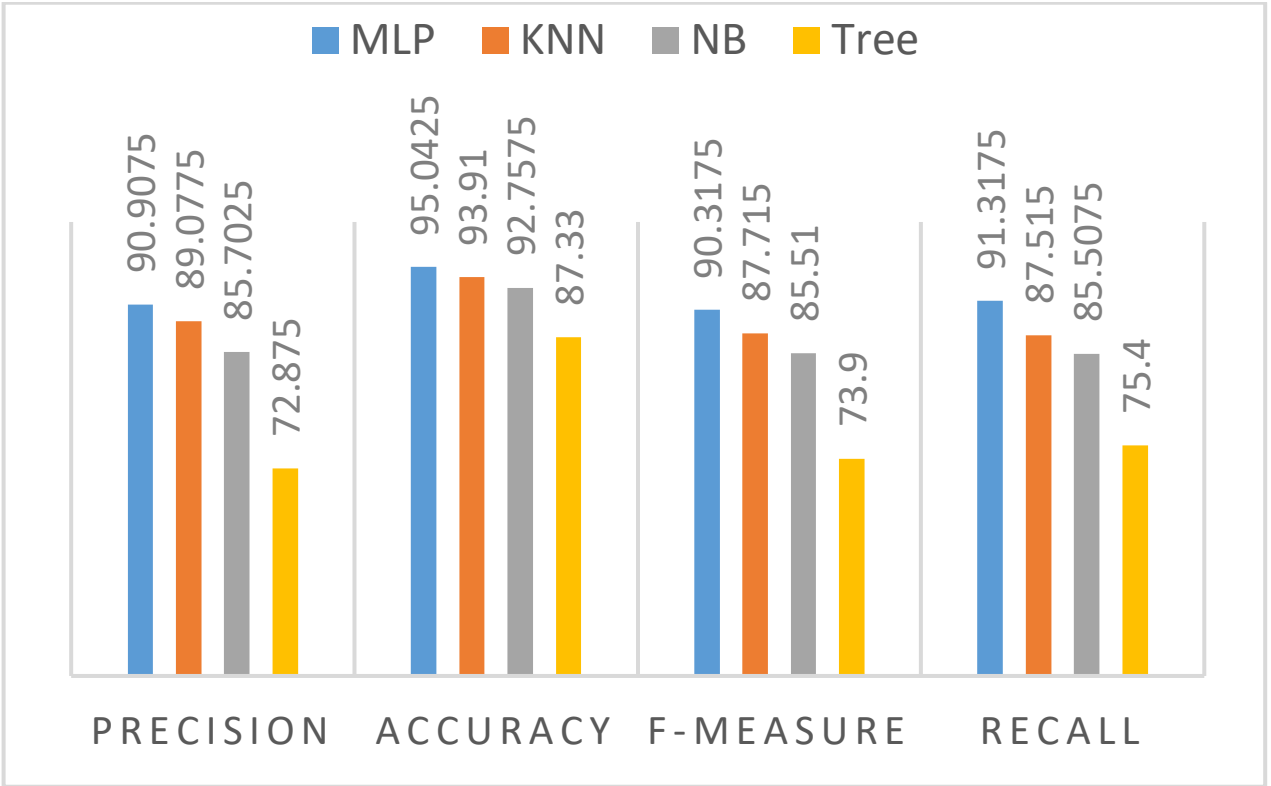

Fig. 7. Comparison between different classification algorithms using MFCC and LPC features

TABLE II

THE WORK THAT DONE IN THE FIELD OF ASSISTANCE TO PEDESTRIANS

| Ref. | approach & tools | weakness | Preferences of our proposed system |
|---|---|---|---|
| [19] | Using infrared sensors that installing on glasses | • Use of additional equipment for pedestrian<br>• Infrared sensors cannot detect differences in the objects which have a very similar temperature range<br>• Infrared sensors are extremely expensive | VP does not need to carry extra equipment.<br>Acoustic sensor has a lower price. |
| [20] | Using camera that installing on bicycle helmet | • Use of additional equipment for pedestrian<br>• Expensive for pedestrian<br>• High cost in process<br>• Lack of privacy | VP does not need to carry extra and expensive equipment.<br>The processing of acoustic data has a lower cost than image data.<br>Preserve privacy. |
| [21] | Using a walking assistant robotic based on computer vision and tactile perception | • Use of additional and very big equipment for pedestrian<br>• High cost in process<br>• Expensive for pedestrian | VP does not need to carry extra and expensive equipment.<br>Preserve privacy.<br>Low cost in processing. |
| [22]<br>[23]<br>[1] | Using vehicle to pedestrian connection | • Get extra information from vehicle driver such as Exact location, speed, and age<br>• Some of this article connect only one car and one pedestrian | Don't need extra information.<br>Only give pedestrian location.<br>Risky vehicle alert send to all of people that placed in danger area (not one person). |

TABLE III

THE WORK THAT DONE IN THE FIELD OF VEHICLE SOUND CLASSIFICATION

| Ref. | classes | accuracy & weakness | Preference of our proposed system |
|---|---|---|---|
| [24] | • Car<br>• Bike<br>• Lorry<br>• Truck | • Low accuracy 86.86% in classification<br>• Lack of direction detection<br>• Don't attention to environment sound | Accuracy is between (93.93%, 96.77%).<br>Direction detection by directional microphone.<br>No vehicle classification includes some of environment. |
| [9] | • car<br>• Bike<br>• Lorry<br>• Truck | • Maximum of neural network training results for the classification of the type of vehicle is 94.5<br>• Not mentioned accuracy of test in neural network<br>• Lack of direction detection | Test accuracy between (93.93%, 96.77%).<br>Direction detection by directional microphone. |
| [3] | • Heavy<br>• Medium<br>• Light<br>• horn | • low accuracy 67.4% in classification<br>• Lack of direction detection | |
| [8] | • Bus vehicle<br>• Car vehicle<br>• Motor vehicle<br>• Truck vehicle | • Low accuracy 83% in classification<br>• Lack of direction detection | |
| [10] | • Buses<br>• Trucks<br>• 4-wheel drive<br>• sedan | • Don't express the classification accuracy<br>• Lack of direction detection<br>• Lack of clarity in helping pedestrians | Helping pedestrian express in detail.<br>Test accuracy between (93.93%, 96.77%).<br>Direction detection by directional microphone. |
| [25] | • Two type of car<br>• Large truck | • Don't express the classification accuracy<br>• Classification with a limited number | Test accuracy between (93.93%, 96.77%).<br>3 classes of vehicle and 1class of no vehicle. |

benchmarks. Based on our best knowledge, there is not any standard benchmark for VP safety problems; therefore, we provide a set of sound recordings in traffic roads. These data are collected from different type of traffic roads and at the different times of day and night. The number of our collected data is 210 sample that 70, 50, 44, and 46 samples related LH, LL, H, and NV, respectively. Training process was done based on cross-validation (6-fold). To show the effectiveness of the proposed system, we use different feature extraction and classification algorithms. In the first, we express results related to frames then express the results of the received sound.

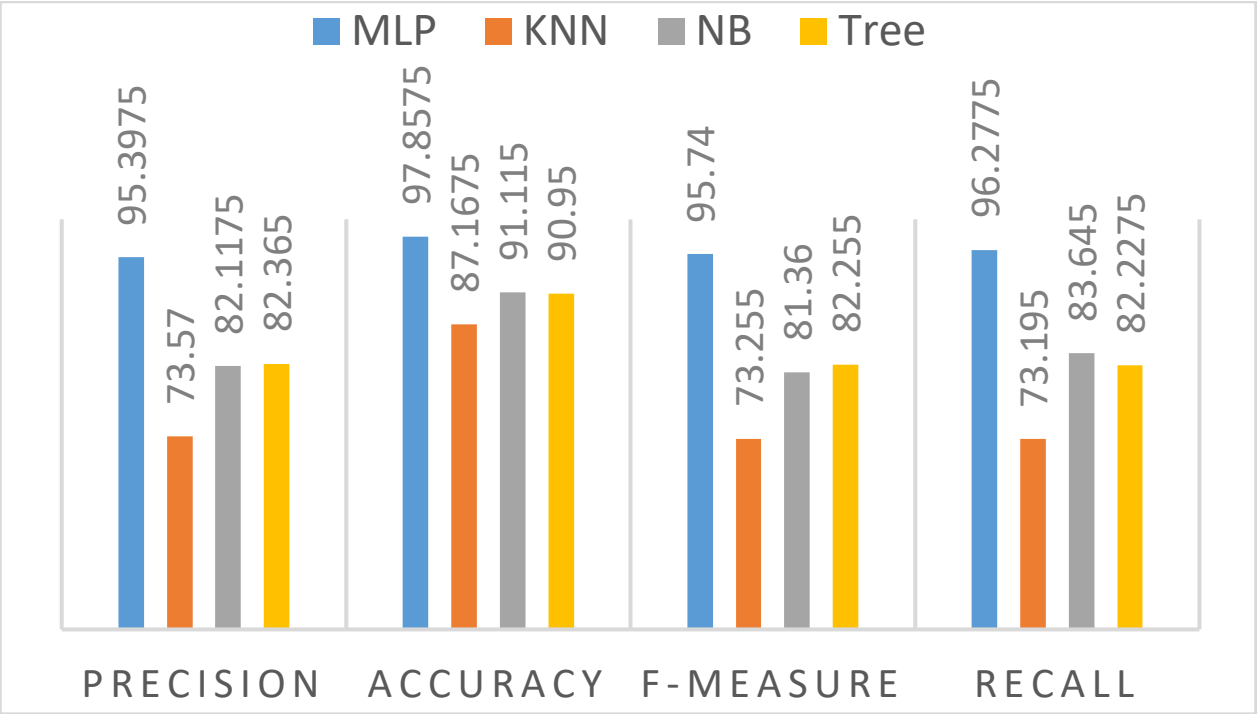

Fig. 8. Comparison between different classification algorithms using all features of Table I

TABLE IV
SENSITIVITY ANALYSIS FOR EIGHT FINAL FRAME

| | H class | LL class | LH class | NV class |
|---|---|---|---|---|
| Precision | 98.30 | 95.87 | 98.38 | 93.47 |
| Recall | 93.54 | 93.93 | 98.38 | 98.85 |
| Accuracy | 98.38 | 96.77 | 99.35 | 97.74 |
| f-measure | 95.86 | 94.89 | 98.38 | 96.08 |

In Fig. 6, the performance of the different classifiers by considering the first five features of Table I, are presented. This figure shows that KNN and MLP on the first five features of Table I provides the best results; but the accuracy in this case is 85.5%. Fig. 7 studies on the effects of MFCC and LPC features. As one can see, the most accuracy corresponds to MLP, which is 95%. Finally, Fig. 8 illustrates the effects of all of the features of Table I. Again, MLP with 97.8% accuracy is the best classifier. Therefore, using all features of Table I, is the best choice in the proposed warning system. According to the different measures including Precision, Recall, Accuracy, and F-measure, one can understand that using MLP classifier algorithm has the best result compared to the other algorithms.

After determining the type of each frame, the proposed system uses eight final frames (unit climax point) for making final decision regarding to the warning. Table IV shows the results of warning with respect to H, LL, LH, and NV classes.

These results reached by voting between in eight final frames. Results given in Table IV, show that the proposed system can classify the different situations between 93.47% and 99.35% that shows that the proposed system is reliable.

## IV. CONCLUSION

In this paper, we proposed a new warning system based on acoustic sensors to warn VP with respect to the risk of vehicles that approach from the behind. The proposed system has the following advantages:

- Only using the acoustic sensors
- Low cost in installing equipment and maintenance in roadside.
- Without necessary to carry extra equipment by VP.
- Efficiency of the system in both of the sunny and the foggy conditions.

In this paper, to help VP in risky conditions, we define some features based on the vehicles sound and we classify them efficiently. We express the new features associate with MFCC and LPC to classifiers to increase the system accuracy. The proposed system can be used efficiency in rainy air if we use a Proper filter. In this paper, the proposed system classifies the received sound in four classes: heavy vehicle, light vehicle with low speed, light vehicle with high speed, and no vehicle with classification accuracy between 93.47% and 98.38%, which shows the applicability of the proposed system in real situations.

## V. REFERENCES

[1] H. R. Eftekhari and M. Ghatee, "An inference engine for smartphones to preprocess data and detect stationary and transportation modes," *Transportation Research Part C: Emerging Technologies*, vol. 69, pp. 313-327, 2016.

[2] M. M. Bejani and M. Ghatee, "A context aware system for driving style evaluation by an ensemble learning on smartphone sensors data," *Transportation Research Part C: Emerging Technologies*, vol. 89, pp. 303-320, 2018.

[3] H. R. Eftekhari and M. Ghatee, "Hybrid of discrete wavelet transform and adaptive neuro fuzzy inference system for overall driving behavior recognition," *Transportation Research Part F: Traffic Psychology and Behaviour*, vol. 58, pp. 782-796, 2018.

[4] H. Eftekhari and M. Ghatee, "A similarity-based neuro-fuzzy modelling for driving behavior recognition applying fusion of smartphone sensors," *Journal of Intelligent Transportation Systems: Technology, Planning, and Operations*, 2019.

[5] R. Bastani Zadeh, M. Ghatee and H. R. Eftekhari, "Three-Phases Smartphone-Based Warning System to Protect Vulnerable Road Users Under Fuzzy Conditions," *IEEE Transactions on Intelligent Transportation Systems*, pp. 1-13, 2017.

[6] J. Gajda, R. Sroka and M. Stencel, "A vehicle classification based on inductive loop detectors," in *Instrumentation and Measurement Technology Conference*, Budapest, Hungary, 2001.

[7] J. George, A. Cyril, B. I. Koshy and L. Mary, "Exploring Sound Signature for Vehicle Detection and Classification Using ANN," *International Journal on Soft Computing (29 -36)*, vol. 4, no. 2, pp. 29-36, 2013.

[8] G. Padmavathi, D. Shanmugapriya and M. Kalaivani, "A Study on Vehicle Detection and Tracking Using Wireless Sensor Networks," *Wireless Sensor Network*, vol. 2, no. 2, pp. 173-185, 2010.

[9] A. Y.Nooralahiyana, M. Doughertyb, D. McKeownc and H. R.Kirbyd, "A field trial of acoustic signature analysis for vehicle classification," *Transportation Research Part C: Emerging Technologies*, vol. 5, no. 3.4, pp. 165-177, 1997.

[10] A. Averbuch, E. Hulata, V. Zheludev and I. Kozlov, "A Wavelet Packet Algorithm for Classification and Detection of Moving Vehicles," *Multidimensional Systems and Signal Processing*, vol. 12, no. 1, pp. 9-31, 2001.

[11] M. V. Ghiurcau and C. Rusu, "Vehicle sound Classification. Application and Low Pass Filtering Influence," in *Signal, Circuits and System*, Lasi, Romania, 2009.

[12] A. D. Mayvana, S. A. Beheshtib and M. H. Masoomc, "Classification of Vehicles Based on Audio Signals using Quadratic Discriminant Analysis and High Energy Feature Vectors," *International Journal on Soft Computing (IJSC)*, vol. 6, no. 1, pp. 53-64, 2015.

[13] M.P.Paulraj, A. H. Adom, S. Sundararaj and N. B. A. Rahim, "Moving Vehicle Recognition and Classification based on Time Domain Approach," *Procedia Engineering*, vol. 53, pp. 405-410, 2013.

[14] J. Lee and A. Rakotonirainy, "Acoustic Hazard Detection for Pedestrians with Obscured Hearing," *IEEE Transactions on Intelligent Transportation Systems*, vol. 12, no. 4, pp. 1640 - 1649, 2011.

[15] N. Lubbe and E. Rosén, "Pedestrian crossing situations: Quantification of comfort boundaries to guide intervention timing," *Accident Analysis & Prevention*, vol. 71, pp. 261-266, 2014.

[16] X. Jiang, W. Wang and K. Bengler, "Intercultural Analyses of Time-to-Collision in Vehicle–Pedestrian Conflict on an Urban Midblock Crosswalk," *IEEE Transactions on Intelligent Transportation Systems*, vol. 16, no. 2, p. 10481053, 2015.

[17] R. San-Segundo, J. M. Montero, R. Barra-Chicote, F. Fernández and J. M. Pardo, "Feature extraction from smartphone inertial signals for human activity segmentation," *Signal Processing*, vol. 120, pp. 359-372, 2016.

[18] N. Desai, P. Dhameliya and P. Desai, "Feature Extraction and Classification Techniques for Speech Recognition: A Review," *International Journal of Emerging Technology and Advanced Engineering*, vol. 3, no. 12, pp. 367-371, 2013.

[19] J. A. Gómez-Tejedor, J. C. Castro-Palacio2 and J. A. Monsoriu, "The acoustic Doppler effect applied to the study of linear motions," *European Journal of Physics*, vol. 35, no. 2, 2014.

[20] J. Scholliersa, D. Bellb, A. Morrisc and A. B. Garcíad, "Improving safety and mobility of Vulnerable Road Users through ITS," in *Transport Research Arena*, paris, 2014.

[21] T. Williams, P. Alves, G. Lachapelle and C. Basnayake, "Evaluation of GPS-based methods of relative positioning for automotive safety applications," *Transportation research part C: emerging technologies*, vol. 23, pp. 98-108, 2012.

[22] E. L. Salomons and P. J. M. Havinga, "A Survey on the Feasibility of Sound Classification on Wireless," *Sensors*, vol. 15, pp. 7462-7498, 2015.

[23] G. Korres, A. E. Issawi and a. M. Eid, "TActile Glasses (TAG) for Obstacle Avoidance," *Springer, Cham*, vol. 8515, pp. 741-749, 2014.

[24] T. Schwarze, M. Lauer, M. Schwaab, M. Romanovas, S. Bo¨hm and T. Ju¨rgensohn, "A Camera-Based Mobility Aid for Visually Impaired People," *KI - Künstliche Intelligenz*, vol. 30, no. 1, pp. 29-36, 2016.

[25] D. Ni, A. Song, L. Tian, X. Xu and D. Chen, "A Walking Assistant Robotic System for the Visually Impaired Based on Computer Vision and Tactile Perception," *International Journal of Social Robotics*, pp. 1-12, 2015.

[26] M. Bagheri, M. Siekkinen and J. K. Nurminen, "Cellular-based Vehicle to Pedestrian (V2P) Adaptive Communication for Collision Avoidance," in *IEEE International Conference on Connected Vehicles and Expo (ICCVE)*, Vienna, Austria , 2014.

[27] L. Zhenyu, P. Lin, Z. Konglin and Z. Lin, "Design and evaluation of V2X communication system for vehicle and pedestrian safety," *The Journal of China Universities of Posts and Telecommunications*, vol. 22, no. 6, pp. 18-26, 2015.

[28] N. A. Rahim, P. MP and A. Adom, "Adaptive Boosting with SVM Classifier for Moving Vehicle Classification," *Procedia Engineering*, vol. 53, pp. 411-419, 2013.

[29] J. Berdnikova, T. Ruuben, V. Kozevnikov and S. Astapov, "Acoustic Noise Pattern Detection and Identification Method in Doppler System," *Electronics and Electrical Engineering*, vol. 18, no. 8, pp. 65-68, 2012.

Biography

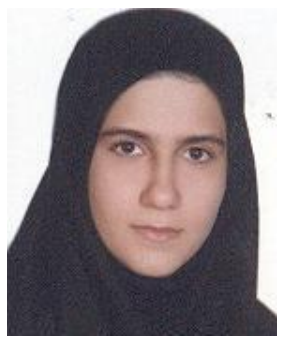

**Masoomeh Khalili is** an MSc student under supervision of Dr. Mehdi Ghatee. She focuses on intelligent transportation systems, vulnerable road user safety and machine learning.

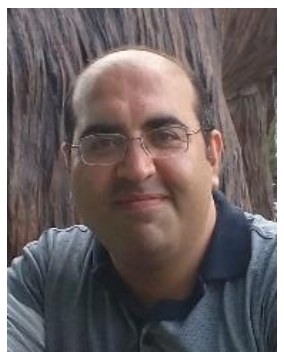

**Mehdi Ghatee** is an Associate Professor with Department of Computer Science, Amirkabir University of Technology, Tehran, IRAN. His major is Intelligent Transportation Systems, Decision Support System and Neural Network. He has written more than 75 papers on national and international journals and conferences.

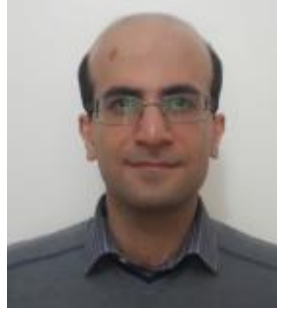

**Mehdi Teimouri,** is an Assistant Professor with Faculty of New Sciences and Technologies, University of Tehran, Tehran, Iran. His major is Electrical Engineering and Signal Processing.

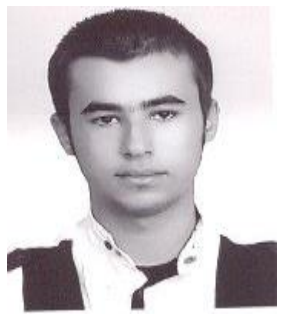

**Mohammad Mahdi Bejani** is currently a PhD student in the Department of Computer Science, Amirkabir University of Technology. His research interests includes intelligent transportation systems, data mining and machine learning.

.